\documentclass[a4paper,12pt]{article}
\usepackage{amsmath, amssymb, bm}
\usepackage{graphicx,color,cite}
\usepackage{xr-hyper}
\usepackage{hyperref}
\usepackage{color}
\usepackage{titlesec}
\usepackage{tocloft}
\usepackage{float}
\usepackage{placeins}
\usepackage{subcaption}
\usepackage{authblk}
\usepackage[margin=2cm]{geometry}

\begin{document}
\title{Dipolar Modeling of Multipolar Metasurfaces}

\author[1]{Hossein Allahverdizadeh} 
\author[1]{Karim Achouri\thanks{karim.achouri@epfl.ch}}

\affil[1]{\small Institute of Electrical and Microengineering\\
École Polytechnique Fédérale de Lausanne (EPFL)\\
Laboratory for Advanced Electromagnetics and Photonics\\
Lausanne, Switzerland}
\date{\today}
\maketitle

\begin{abstract}
Multipolar decomposition is a powerful tool for analyzing and designing metasurfaces, but its practical application is often limited by the mathematical complexity that arises when a large number of multipole moments must be taken in to account. To minimize this modeling complexity without sacrificing accuracy, we present an efficient method that exploits the coordinates origin dependence of spherical multipole moments. We show that the optimal origins for minimizing higher-order contributions, such as quadrupoles and octupoles, depend strictly on the spatial parity of the electromagnetic response. This is achieved by modeling a metasurface response using multipolar generalized sheet transition conditions (GSTCs). By separating the GSTCs into independent even and odd parity components, we can evaluate the electric and magnetic discontinuities at distinct physical positions. This parity-splitting framework allows us to systematically suppress unwanted higher-order terms and reconstruct the complete scattering parameters using only the dipole moments. We validate our analytical approach using two numerical examples: vertically asymmetric dielectric cones on a substrate, and a horizontally symmetry-broken metasurface supporting a double quasi-bound state in the continuum resonance. In both cases, the retrieved scattering parameters show excellent agreement with full-wave simulations. This method provides a simple, physically intuitive framework that simplifies the modeling of geometrically complex and non-local metasurfaces down to a purely dipolar level.
\end{abstract}

\section{Introduction}

A multipolar approach offers a clear physical picture of the scattering mechanisms responsible for advanced metasurface functionalities. These include the generalized Kerker effect~\cite{liu2018generalized}, symmetry-protected or accidental bound states in the continuum (BICs)~\cite{koshelev2018asymmetric}, the non-radiating anapole state~\cite{doi:10.1021/acs.jpcc.6c00597}, as well as applications like wavefront shaping, metalenses, and perfect absorption~\cite{babicheva2024mie}. 

The practical use of multipolar expansions requires a careful control over the number of multipole moments that are considered. If too few multipole moments are taken into account, then the accuracy of the modeling approach may not be sufficient to fully capture all scattering features. Whereas, too many multipole moments, increases the modeling accuracy but comes at the cost of significantly more complexity and heavy computation. This aspect is fundamental to multipolar metasurface modeling and design.

Several factors govern how many multipole moments are required to accurately capture the response of a metasurface, including the electrical size, geometry and material parameters of the unit cell~\cite{alaee_multipoles_2018}, lattice couplings~\cite{rahimzadegan2022comprehensive}, the spatial symmetries of the structure~\cite{sadrieva2019multipolar,achouri2023a}, the incident angle~\cite{Allahverdizadeh}, and the selected origin of the system of coordinates used to perform the multipolar decomposition~\cite{kildishev2025art}. 

Because these factors often require a large number of multipole moments, theoretical modeling can quickly become computationally very intensive. For instance, a common approach for modeling metasurfaces consists in representing them as zero-thickness sheets of homogeneous currents capable of generating field discontinuities~\cite{idemen2011,kuester2003,holloway2012,achouri2021electromagnetic}. This framework relies on generalized sheet transition conditions (GSTCs) and has been recently extended to include the contributions of multipole moments up to the octupolar regime~\cite{achouri2022b, tiukuvaara2023,Allahverdizadeh}. As such, it is the modeling framework that shall be used in this work. One important feature of this GSTCs-based metasurface model is that it allows the effective susceptibilities of a metasurface to be directly retrieved from the incident and scattered electromagnetic fields. However, retrieving the effective multipolar susceptibilities rapidly becomes challenging, since accurately capturing the full response of a metasurface often requires several hundreds of susceptibility components~\cite{achouri2022b, tiukuvaara2023}. Consequently, restricting the model to a purely dipolar description is therefore highly desirable to keep the problem manageable.

Similarly, the T-matrix method, which is also used to model the scattering properties of meta-atoms and their collective lattice interactions~\cite{zerulla2023matrix}, faces a similar bottleneck. The T-matrix relates the incident and scattered fields using a multipolar expansion. However, when high-order multipole moments must be included to achieve accurate convergence, the size of the matrix increases drastically. This makes both the analytical formulation and the numerical calculations much more complicated. In this context, reducing the electromagnetic response strictly to a dipolar level would also offer a significant advantage~\cite{gladyshev2023inverse}. A purely dipolar T-matrix would limit the system to a minimal size and simplifies the matrix equations, which is highly beneficial for implementing inverse design approaches and accelerating metasurface optimization. Therefore, a method that systematically suppresses higher-order contributions to preserve a pure dipolar description is very useful across all of these modeling frameworks.

In this work, we use the multipolar GSTCs-based metasurface model developed in~\cite{Allahverdizadeh}. Our objective is to preserve computational accuracy while minimizing the number of required multipole moments to reduce the modeling complexity and computational cost of the approach. To this end, we exploit the inherent dependence of multipole moments on the choice of coordinate-system origin used in the multipolar decomposition~\cite{kildishev2025art}. By appropriately translating the origin, the amplitudes of higher-order multipole moments can be significantly reduced, potentially allowing the metasurface to be accurately modeled using only dipole moments. However, as pointed out in~\cite{kildishev2025art}, the amplitude of the multipole moments with different spatial parities typically change differently under a translation of the origin. Consequently, the contributions of multipole moments with both parities cannot, in general, be minimized simultaneously using a single origin. To address this issue, we decompose the GSTCs into two distinct parity-dependent relations, allowing each relation to be associated with the optimal origin for the corresponding parity of multipole moments. As a result, the proposed approach allows us to accurately model metasurfaces that would typically require several orders of multipole moments using only electric and magnetic dipole moments.

\section{Theoretical Background}

\subsection{Conventional Metasurface Modeling}

We model a metasurface as a zero-thickness sheet of induced effective electric and magnetic current densities, $\mathbf{J}_\text{s}$ and $\mathbf{K}_\text{s}$, respectively. Using Maxwell equations in the frequency domain and assuming the time dependence $e^{j\omega t}$, the electromagnetic fields that interact with a metasurface at $z=0$ must satisfy
\begin{subequations}
\label{eq_maxwell}
\begin{align}
    \nabla \times \mathbf{H} &= +\mathbf{J}_\text{s} \delta(z) + j \omega \epsilon_0 \mathbf{E}, \label{eq:maxwell_H} \\
    \nabla \times \mathbf{E} &= -\mathbf{K}_\text{s} \delta(z) - j \omega \mu_0 \mathbf{H}, \label{eq:maxwell_E}
\end{align}
\end{subequations}
where $\delta(z)$ denotes the Dirac delta distribution. 

Equations~\eqref{eq_maxwell} are valid everywhere in space around the metasurface but are difficult to evaluate at the metasurface itself due to the discontinuities introduced by the Dirac deltas. To overcome this issue, we next concentrate our attention only on the fields and currents that are directly at the metasurface itself. In other words, we have to transform~\eqref{eq_maxwell} into a set of boundary conditions.

To do so, we decompose the curl operators into their transverse ($\nabla_{\text{t}}$) and normal ($\hat{\mathbf{n}}\partial_z$) components. Then, we integrate~\eqref{eq_maxwell} along the $z$-axis from $z = 0^-$ to $z = 0^+$. This integration transforms the terms involving the partial derivative $\partial_z$ into the field jump $\Delta \mathbf{\Psi} = \mathbf{\Psi}(0^+) - \mathbf{\Psi}(0^-)$, where $\mathbf{\Psi} = \{\mathbf{E}, \mathbf{H}\}$. As a consequence, this connects the normal components of the surface currents ($J_n$ and $K_n$) to the tangential field discontinuities through the transverse divergence. By retaining only the tangential components, we obtain the Generalized Sheet Transition Conditions (GSTCs)~\cite{idemen2011,kuester2003,albooyeh2016electromagnetic,achouri2021electromagnetic}
\begin{subequations}
\label{eq_GSTCs}
\begin{align}
\hat{\mathbf{z}} \times \Delta \mathbf{H}_{\text{t}} &= + \mathbf{J}_{\text{t}} + \frac{1}{j \omega \mu_0} \nabla_{\text{t}} \times (\hat{\mathbf{n}} K_n), \label{eq:GSTCs_H} \\
\hat{\mathbf{z}} \times \Delta \mathbf{E}_{\text{t}} &= -\mathbf{K}_{\text{t}} + \frac{1}{j \omega \epsilon_0} \nabla_{\text{t}} \times (\hat{\mathbf{n}} J_n). \label{eq:GSTCs_E}
\end{align}
\end{subequations}

In the remainder of this paper, we will regularly decompose the response of a metasurface in terms of the spatial parity of the components that constitute its overall electromagnetic response. We therefore emphasize here that the electric and magnetic fields are, respectively, polar (odd under parity) and axial (even under parity) vectors~\cite{jackson2012classical}. This distinction allows us to determine the parity of all the terms appearing in~\eqref{eq_GSTCs}, which are summarized in Table~\ref{tab:parity_GSTCs}.

\begin{table}[H]
\centering
\caption{Parity of electromagnetic fields and current densities in the GSTCs.}
\label{tab:parity_GSTCs}
\begin{tabular}{@{}lcc@{}}
\hline
\textbf{Quantity} & \textbf{Symbol} & \textbf{Parity} \\ \hline
Electric Field & $\mathbf{E}$ & Odd ($-$) \\
Magnetic Field & $\mathbf{H}$ & Even ($+$) \\ \hline
Tangential Electric Current & $\mathbf{J}_{\text{t}}$ & Odd ($-$) \\
Tangential Magnetic Current & $\mathbf{K}_{\text{t}}$ & Even ($+$) \\ \hline
Normal Electric Current & $J_n$ & Odd ($-$) \\
Normal Magnetic Current & $K_n$ & Even ($+$) \\ \hline
Electric Field Jump & $\Delta \mathbf{E}_{\text{t}}$ & Even ($+$) \\
Magnetic Field Jump & $\Delta \mathbf{H}_{\text{t}}$ & Odd ($-$) \\ \hline
\end{tabular}
\end{table}

\subsection{Multipolar Decomposition of Equivalent Current}

To rigorously model the scattering response of a metasurface with geometrically complex, and especially electrically large, unit cells, we must extend the GSTCs to include multipole moments, as discussed in~\cite{achouri2022b,tiukuvaara2023,Allahverdizadeh}. To do so, we start by considering that the scattering response of each scattering particle in the metasurface array may be captured by the volumetric electric current density, $\mathbf{J}(\mathbf{r})$, inside it. This allows us to define the corresponding electric vector potential, $\mathbf{A}(\mathbf{r})$, as~\cite{jackson2012classical}
\begin{equation}
\label{eq:vector_potential}
\mathbf{A}(\mathbf{r}) = \frac{\mu}{4\pi} \int_{V} \mathbf{J}(\mathbf{r'}) \frac{e^{-jkR}}{R} \, dV', 
\end{equation}
Now, we model the scattering particles as point-like sources that exhibit the homogeneous equivalent current $\mathbf{J}_{\text{eq}}$. This reduces~\eqref{eq:vector_potential} into
\begin{equation}
\label{eq:A_homogenized}
\mathbf{A}(\mathbf{r}) = \frac{\mu}{4\pi} \mathbf{J}_{\text{eq}} \frac{e^{-jkr}}{r}.
\end{equation}
The equivalent current $\mathbf{J}_{\text{eq}}$ is now expanded into its multipolar components as~\cite{groot1972,simovski2018,jackson2012classical,nanz2016toroidal}
\begin{equation}
\label{eq_curent}
    \mathbf{J_{\text{eq}}}(\mathbf{r}) = j\omega\mathbf{p} - \frac{j\omega}{2} \mathbf{Q}^\text{(e)} \cdot \nabla + \nabla \times \mathbf{m} + \frac{j\omega}{6} \mathbf{O}^\text{(e)} : \nabla\nabla - \frac{1}{2} \nabla \times (\mathbf{Q}^\text{(m)} \cdot \nabla) + \dots,
\end{equation}
where the name of each multipole moment is given in Table~\ref{tab:multipole_parity_final}. Note that these are \emph{spherical} multipoles, which are defined in the SI from~\cite{nanz2016toroidal,alaee_multipoles_2018}. These multipoles moments are therefore symmetric and traceless and automatically include all toroidal contributions.

Equation~\eqref{eq_curent} forms the basis of the multipolar extension of the GSTCs developed in~\cite{Allahverdizadeh} and which we shall briefly review in the next section. Before delving into this, we note that the parity of the multipole moments is determined by their transformation properties under the spatial inversion operator $\mathcal{P}: \mathbf{r} \to -\mathbf{r}$. Given that the electric current density is odd under parity, as given in Table~\ref{tab:parity_GSTCs}, the individual multipolar contributions must satisfy specific symmetry requirements to ensure the total current maintains this property. The resulting parity of the electric and magnetic multipole moments are summarized in Table~\ref{tab:multipole_parity_final}.
\begin{table}[H]
\centering
\caption{Parity of Electric and Magnetic Multipole Moments}
\label{tab:multipole_parity_final}
\begin{tabular}{@{}lcc@{}}
\hline
\textbf{Multipole Moment} & \textbf{Symbol} & \textbf{Parity} \\ \hline
Electric Dipole           & $\mathbf{p}$    & Odd ($-$)       \\
Magnetic Dipole           & $\mathbf{m}$    & Even ($+$)      \\
Electric Quadrupole       & $\mathbf{Q}^{(\text{e})}$ & Even ($+$)      \\
Magnetic Quadrupole       & $\mathbf{Q}^{(\text{m})}$ & Odd ($-$)       \\
Electric Octupole         & $\mathbf{O}^{(\text{e})}$ & Odd ($-$)       \\
Magnetic Octupole         & $\mathbf{O}^{(\text{m})}$ & Odd ($-$)       \\ \hline
\end{tabular}
\end{table}

\subsection{Multipolar Extension of the GSTCs}

In~\eqref{eq_curent}, we have constructed an equivalent surface current of the metasurface scattering particles based on the multipolar moment expansion. Now, to obtain the multipolar GSTCs, our next goal is to calculate the far-field radiation produced by an infinite homogeneous sheet that harbors a normalized version of this current. This will then allow us to determine the field discontinuities across this sheet, which represents the metasurface, and ultimately derive multipolar GSTCs. The fields radiated by an infinite homogeneous sheet are given by~\cite{Allahverdizadeh} 
\begin{subequations}\label{field_potential}
\begin{align}
\mathbf{E}&= -\frac{1}{2k_z\omega\epsilon} [\boldsymbol{\nabla}\boldsymbol{\nabla} + k^2\overline{\overline{\mathbf{I}}}] \cdot \tilde{\mathbf{J}}_\text{eq} e^{-jk_z|z|}, \\[1.5ex]
\mathbf{H} &= \frac{1}{2jk_z} \boldsymbol{\nabla} \times \tilde{\mathbf{J}}_\text{eq} e^{-jk_z|z|},
\end{align}
\end{subequations}
where $\tilde{\mathbf{J}}_\text{eq} = \mathbf{J}_\text{eq}/S$ with $S$ being the surface area of the metasurface unit cell.

By substituting the multipolar expansion of the equivalent current~\eqref{eq_curent} into~\eqref{field_potential} and solving the resulting equations for the fields at $z=0^+$ and $z=0^-$, we directly obtain the multipolar GSTCs relations~\eqref{eq_deltaE_parallel} and~\eqref{eq:deltaH_parallel} given in the SI. For the specific case of TE-polarized oblique wave propagation, the GSTCs reduce to a set of relations involving the $y$-component of the electric field jump and the $x$-component of the magnetic field jump~\cite{Allahverdizadeh}
\begin{subequations}
\label{psi_TE_EH}
\begin{align}
    \Delta E_x^{\text{TE}} &= \tilde{M}_x + \tilde{Q}_{yz}^{\text{(e)}} + \tilde{Q}_{xx}^{\text{(e)}} + \tilde{Q}_{zz}^{\text{(e)}} + \tilde{O}_{xzy}^{\text{(e)}} + \tilde{O}_{yzz}^{\text{(m)}} , \label{psi_TE_E} \\
    \Delta H_y^{\text{TE}} &= \tilde{P}_y + \tilde{M}_z + \tilde{Q}_{xz}^{\text{(m)}} + \tilde{Q}_{yx}^{\text{(m)}} + \tilde{O}_{yxx}^{\text{(e)}} + \tilde{O}_{xzz}^{\text{(e)}}, \label{psi_TE_H}
\end{align}
\end{subequations}

where $\tilde{P}$, $\tilde{M}$, $\tilde{Q}$, and $\tilde{O}$ represent normalized versions of the electric dipole, magnetic dipole, quadrupole, and octupole moments respectively given in Table~\ref{tab:te_multipoles} of the SI. Note that some of these normalized moments include the wavevector components $k_x$ and/or $k_z$. Since these components are odd under parity, we provide an updated version of Table~\ref{tab:multipole_parity_final} in Table~\ref{tab:te_multipoles} in the SI that properly accounts for the parity of these normalized moments. It follows that the GSTCs in~\eqref{psi_TE_EH} are split into two equations that precisely correspond to the parity of each of the involved quantities. Specifically, the boundary condition for ${\Delta E}$ is even, whereas the one for ${\Delta H}$ is odd under parity.

Equation~\ref{psi_TE_EH} includes multipole moments up to the octupolar terms. While this expansion can be extended to even higher orders if necessary, implementing a large number of multipoles increases the modeling complexity and makes the design process much harder, despite improving precision. This raises an important question: is it possible to model the metasurface response by truncating the expansion to a lower number of multipoles to avoid a complex modeling process, without sacrificing precision?

To answer this question, we consider a fundamental aspect of multipolar theory, which is that the multipole moments are dependent on the origin of the system of coordinates used to perform the multipolar decomposition. This dependency is clearly visible in the presence of the position vector $\mathbf{r}$ in the integrals of Table~\ref{tab:multipoles} in the SI. This means that the amplitude of the multipole moments changes when considering different origins, which therefore suggests that we may possibly find an optimal origin for the multipolar decomposition that minimizes the strength of the higher-order multipoles. We will see in the remainder of this work that it is indeed possible to minimize the contribution of higher-order multipoles but this usually requires different origins for the even and odd parts of the GSTCs in~\eqref{psi_TE_EH}, which is similar to the concept recently discussed in~\cite{kildishev2025art} for single particles. In the next section, we will show how to properly split the GSTCs to minimize the complexity of the multipolar modeling of metasurfaces.

\subsection{Modeling of a Substrated Metasurface}

We now show how the conventional multipolar metasurface modeling framework, developed in~\cite{Allahverdizadeh}, is typically applied to model the scattering response of a metasurface that is surrounded by different substrate and superstrate. We will then show how the complexity of this multipolar model may be reduced by translating the origin of the system of coordinates and effectively removing the contributions from higher-order multipoles. We shall start by discussing the placement of the GSTCs along the $z$-direction (normal to the metasurface) and will discuss the case of in-plane translations in Sec.~\ref{sec_inplane}.

We consider a practical scenario where a metasurface scattering particles possess a finite physical thickness $d$ and are supported by a dielectric substrate. This metasurface is modeled as a zero-thickness sheet, using the GSTCs in~\eqref{psi_TE_EH}, placed at an arbitrary vertical position, $z$, where the multipolar decomposition will be performed. The corresponding metasurface reflection and transmission coefficients, at the position of the GSTCs, are given by $r_{\text{ms}}(z)$ and $t_{\text{ms}}(z)$, respectively. We start by considering a simple case where the metasurface lies in a homogeneous background medium and will later include the presence of the substrate. Using~\eqref{psi_TE_EH}, we have that~\cite{Allahverdizadeh}
\begin{subequations}\label{eq:ms_intrinsic}
\begin{align}
    t_{\text{ms}}(z) &= \widehat{\Delta H}(z) + \widehat{\Delta E}(z) + E_\text{i}e^{jk z}, \label{eq:T_ms}\\
    r_{\text{ms}}(z) &= \widehat{\Delta H}(z) - \widehat{\Delta E}(z), \label{eq:R_ms} 
\end{align}
\end{subequations}
where $\widehat{\Delta E}(z) = \Delta E_x^{\text{TE}}(z)/2$, $\widehat{\Delta H}(z) = \eta \Delta H_y^{\text{TE}}(z)/2$ with $\eta$ being the impedance of the background medium and $E_\text{i}$ denotes the normalized ($|E_\text{i}|=1$) complex amplitude of the incident wave at the position of the GSTCs.

\begin{figure}[!h]
    \centering
    \includegraphics[width=0.9\textwidth]{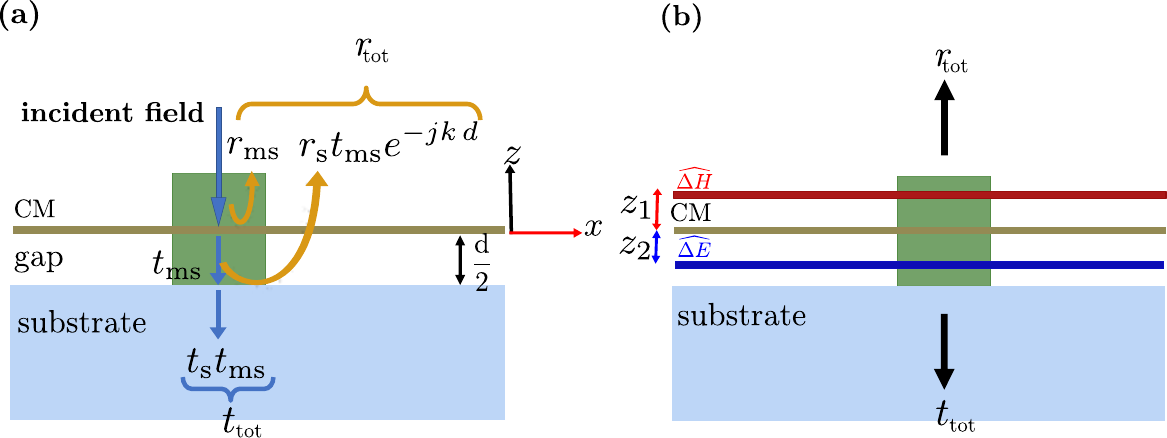}
    \caption{Modeling of a metasurface unit cell on top of a substrate using the GSTCs framework. The scattering particle, in green, is initially replaced by an equivalent zero-thickness sheet (GSTCs) passing through its center of mass (CM). (a) Construction of the total reflection and transmission coefficients. (b) Parity splitting of the GSTCs positions with respect to the CM.}
    \label{fig_5:design}
\end{figure}

In the presence of a substrate, we obtain the total metasurface scattering parameters, $r_\text{tot}$ and $t_\text{tot}$, by taking into account the propagation phase through the thickness of the unit cell and the reflection at the substrate interface. Assuming that the center of the object is located at the center of mass (CM), i.e., $d/2$ for a vertically symmetric particle, the total transmission and reflection are expressed as~\cite{Allahverdizadeh}
\begin{subequations}\label{eq:total_RT_substrate}
\begin{align}
    t_\text{tot} &= t_\text{s} t_{\text{ms}}(z) e^{-jk\left(\frac{d}{2} - z\right)}, \label{eq:T_total} \\[2ex]
    r_\text{tot} &= \left[ r_{\text{ms}}(z) + r_\text{s} t_{\text{ms}}(z) e^{-jk d} \right] e^{-jk z}, \label{eq:R_total}
\end{align}
\end{subequations}

where $r_\text{s}$ and $t_s$ denote the TE-polarized Fresnel reflection and transmission coefficients at the substrate interface, respectively~\cite{novotny2012principles}, and $k$ is the wavenumber in the homogeneous background medium. In~\eqref{eq:total_RT_substrate}, $t_{\text{ms}}$ and $r_{\text{ms}}$ denote the metasurface scattering parameters and are those given in~\eqref{eq:ms_intrinsic} since the GSTCs are placed in the center of mass of the object and thus lie in a homogeneous medium. This is valid provided that the multipole moments that define the field discontinuities in~\eqref{eq:ms_intrinsic} have been obtained in the presence of the substrate from the self-consistent fields~\cite{Allahverdizadeh}. A schematic depicting these different scattering parameters is shown in Fig.~\ref{fig_5:design}a.

As can be seen from Fig.~\ref{fig_5:design}a, the total transmission $t_\text{tot}$ simply incorporates the propagation phase delay $e^{-jk(d/2 - z)}$ from the reference origin $z$ to the physical exit interface at $d/2$. Similarly, the total reflection $r_\text{tot}$ is composed of the direct metasurface reflection ($r_{\text{ms}}$) with a secondary reflection term from substrate ($r_\text{s} t_{\text{ms}} e^{-jk d}$). This second term arises from the forward transmitted wave incident on the substrate interface ($r_\text{s}$) and traversing the element thickness ($d$) leading to a phase accumulation $e^{-jk d}$.

With~\eqref{eq:total_RT_substrate}, we are able to fully capture the scattering response of a substrated metasurface assuming that both GSTCs relations are at the origin, $z$. To exploit the translational dependence of the multipole moments for minimizing higher-order multipole contributions, it is now necessary to establish how the GSTCs evolve under a coordinate shift. 

To do so, we consider the fact that the total scattering parameters in~\eqref{eq:total_RT_substrate} represent far-field quantities that should be independent of the physical position of the GSTCs or that of the system of coordinates used to perform the multipolar decomposition. We now exploit this property to re-express the field discontinuities in~\eqref{eq:ms_intrinsic}, and thus the GSTCs, in terms of the invariant total scattering parameters in~\eqref{eq:total_RT_substrate}.

By substituting~\eqref{eq:ms_intrinsic} into~\eqref{eq:total_RT_substrate} and solving the resulting equations for the field discontinuities, we obtain
\begin{subequations}\label{eq:extracted_jumps_substrate}
\begin{align}
    \widehat{\Delta E}(z) &= \frac{1}{2} \left[ \left( t_\text{tot} - e^{-jk d/2} \right) e^{jk\left(\frac{d}{2} + z\right)} - \left( r_\text{tot} + r_\text{s} t_\text{tot} e^{-jk d/2} \right) e^{-jk z} \right],\label{eq:extracted_E_sub}\\[2ex]
    \widehat{\Delta H}(z) &= \frac{1}{2} \left[ \left( t_\text{tot} - e^{-jk d/2} \right) e^{jk\left(\frac{d}{2} + z\right)} + \left( r_\text{tot} + r_\text{s} t_\text{tot} e^{-jk d/2} \right) e^{-jk z} \right].
    \label{eq:extracted_H_sub} 
\end{align}
\end{subequations}

While~\eqref{eq:extracted_jumps_substrate} explicitly define the GSTCs using invariant scattering parameters and phase varying terms, we can alternatively formulate these expressions using the field discontinuities obtained at a fixed reference origin. By substituting $t_{\text{tot}}$ and $r_{\text{tot}}$ from an arbitrary reference point (see Sec.~\ref{sec:origin_translation} in the SI) chosen here to be the CM at $z=0$, we rewrite our $z$-dependent GSTCs as 
\begin{equation}\label{eq:parity_rotation}
    \begin{bmatrix}
        \widehat{\Delta E}(z)\\[1ex]
        \widehat{\Delta H}(z) 
    \end{bmatrix}
    =
    \begin{bmatrix}
        \cos(k z) & j\sin(k z) \\
        j\sin(k z) & \cos(k z)
    \end{bmatrix}
    \begin{bmatrix}
        \widehat{\Delta E}(0)\\[1ex]
        \widehat{\Delta H}(0) 
    \end{bmatrix},
\end{equation}
where $\widehat{\Delta H}(0)$ and $\widehat{\Delta E}(0)$ correspond to the field discontinuities at the CM. These relations demonstrate that moving the GSTCs along $z$ effectively ``mixes'' the even and odd parity multipole moments calculated at the reference point.

Note that in the special case where the metasurface is embedded in a homogeneous background media (no substrate), relations~\eqref{eq:total_RT_substrate} and~\eqref{eq:extracted_jumps_substrate} may be simplified by setting $d=r_\text{s}=0$ and $t_\text{s}=1$.

\section{Multipolar Minimization via Origin Optimization}

\subsection{Scattering Parameters for Split GSTCs}

An important property of~\eqref{eq:extracted_jumps_substrate} is that the two field discontinuities $\widehat{\Delta E}$ and $\widehat{\Delta H}$ do not need to be evaluated simultaneously at the same $z$-position since they are both independently related to $z$-invariant quantities. This is crucial since we know from~\cite{kildishev2025art} that the optimal position for performing a multipolar decomposition, in the case of a single particle, may not be the same for even and odd multipoles. This is also true for the case of a metasurface, as we shall soon demonstrate. 

In our case, we know from~\eqref{psi_TE_EH} that $\widehat{\Delta E}$ and $\widehat{\Delta H}$ correspond to even and odd parties and are thus related to even and odd multipoles, respectively. This means that in order to reduce the contribution from higher-order even and odd multipoles, we now have to find a way to express the total scattering parameters in terms $\widehat{\Delta E}(z_1)$ and $\widehat{\Delta H}(z_2)$, where $z_1$ and $z_2$ are two different positions along $z$, as illustrated in Figure~\ref{fig_5:design}b.

To retrieve the total scattering parameters from these two different positions, we isolate the transmission and reflection coefficients from~\eqref{eq:extracted_jumps_substrate}. This allows us to construct the linear system of equations given by
\begin{equation}\label{eq:matrix_substrate_sep}
    \begin{bmatrix}
        e^{jk(d/2 + z_1)} & -e^{-jk z_1} \\
        e^{jk(d/2 + z_2)} & e^{-jk z_2}
    \end{bmatrix}
    \begin{bmatrix}
        t_\text{tot} - e^{-jk d/2} \\
        r_\text{tot} + r_\text{s} t_\text{tot} e^{-jk d/2}
    \end{bmatrix}=2
        \begin{bmatrix}
        \widehat{\Delta E}(z_1) \\
        \widehat{\Delta H}(z_2)
    \end{bmatrix}.
\end{equation}
We now solve this system for the total transmission and reflection coefficients to obtain
\begin{subequations}\label{eq:RT_final_substrate}
\begin{align}
    t_\text{tot} &= e^{-jk d/2} \left[ 1 + \frac{\widehat{\Delta E}(z_1) e^{-jk z_2} + \widehat{\Delta H}(z_2) e^{-jk z_1}}{\cos[k(z_2-z_1)]} \right], \label{eq:T_final_sub} \\[2ex]
    r_\text{tot} &= \frac{\widehat{\Delta H}(z_2) e^{jk z_1} - \widehat{\Delta E}(z_1) e^{jk z_2}}{\cos[k(z_2-z_1)]} - r_\text{s} t_\text{tot} e^{-jk d/2}. \label{eq:R_final_sub}
\end{align}
\end{subequations}
  
\begin{figure}[h!]
    \centering
    \includegraphics[width=\textwidth]{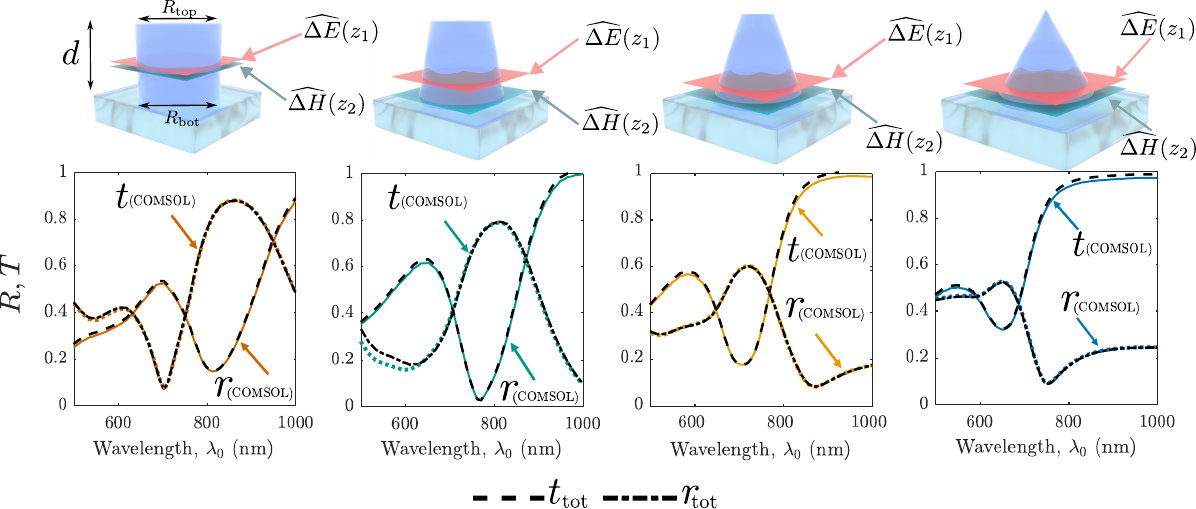}
    \caption{Modeling of the scattering response of four amorphous silicon cones on top of a glass substrate on square lattice with periodicity of $P=300~\text{nm}$. The bottom radius is same for four of them and defines as $R_{\text{bot}}=140~\text{nm}$ and we choose the top radius from the list as $R_{\text{top}}=[R_{\text{bot}},2R_{\text{bot}}/3,R_{\text{bot}}/3,0]$. For each cone, the GSTCs are split and placed at positions $z_1$ and $z_2$ that correspond to where the quadrupolar contributions are minimized. At the bottom, the plots compare the COMSOL simulated scattering parameters (solid lines) to those obtained via our approach (dashed and dotted-dashed lines).}
    \label{fig_2:design}
\end{figure}

In what follows, we consider an example to illustrate and validate our modeling approach based on relations~\eqref{eq:RT_final_substrate}. Specifically, we consider the case of four dielectric cones with different levels of asymmetry, which we refer to as ``conicity'', and controlled by the ratio $R_\text{top}/R_\text{bot}$ where $R_\text{top}$ and $R_\text{bot}$ correspond to the top and bottom radii of the cones, respectively. The cones have a height $d$, they are made of amorphous silicon and stand on a glass substrate, as shown at the top of Fig.~\ref{fig_2:design}.

For each cone, we identify two positions at which the electric and magnetic quadrupolar contributions are minimized. The exact optimal positions that suppress all higher-order multipolar contributions, leaving only the dipolar response, will be discussed in later sections. For each degree of conicity, Fig.~\ref{fig_2:design} compares the scattering parameters obtained numerically using COMSOL with those retrieved from~\eqref{eq:RT_final_substrate}. To determine $r_{\text{tot}}$ and $t_{\text{tot}}$ from these expressions, we first compute the multipole moments at the optimal positions, $z_1$ and $z_2$, using the integrals listed in Table~\ref{tab:multipoles} of the SI (see also Sec.~\ref{sec:Multipole_conicity} for additional details). Next, the corresponding field discontinuities are obtained from~\eqref{psi_TE_EH}. Finally, substituting these quantities into~\eqref{eq:RT_final_substrate} yields the total reflection and transmission coefficients, $r_{\text{tot}}$ and $t_{\text{tot}}$.

Additional details regarding the multipolar responses of these structures are provided in Sec.~\ref{sec:Multipole_conicity} of the SI. In particular, the diagrams in Fig.~\ref{fig_10:design} show that, as the degree of asymmetry increases, the multipole moments become increasingly sensitive to translations of the origin. As a result, highly asymmetric structures exhibit more pronounced minima in their higher-order multipolar contributions compared to more symmetric geometries. Moreover, Fig.~\ref{fig_9:design} illustrates the evolution of the multipole moments for the cone with the highest degree of conicity ($R_\text{top}=0$) as the origin is displaced from its center. Finally, Sec.~\ref{sec:Erroe} of the SI discusses the situations in which the expressions in~\eqref{eq:RT_final_substrate} become nearly divergent, namely when the denominator $\cos[k(z_2-z_1)]$ approaches zero.

\subsection{Multipolar Minimization via Out-of-Plane Origin Translation}

In the following sections, we aim to identify the optimal choice of origin for which the scattering response of the metasurface can be fully described using only electric and magnetic dipole moments, while all higher-order multipolar contributions are minimized.

This may be achieved by comparing the amplitudes of the multipole moments evaluated at different positions along the $z$ axis with those obtained at a reference point, chosen here as the center of mass (CM) located at $z=0$. To this end, we use relations~\eqref{eq:parity_rotation}, which describe how the GSTCs transform under translations away from a reference origin. In these expressions, the field discontinuities $\widehat{\Delta E}(0)$ and $\widehat{\Delta H}(0)$ correspond, through~\eqref{psi_TE_EH}, to the multipole moments evaluated at the CM, whereas $\widehat{\Delta E}(z)$ and $\widehat{\Delta H}(z)$ denote the corresponding quantities evaluated at an arbitrary position $z$. Combining these two relations yields
\begin{subequations}\label{eq:conventional_expansion}
\begin{align}
    \frac{1}{2}\Bigl[&\tilde{M}_y(z_1) + \tilde{Q}_{xz}^{\text{(e)}}(z_1) + \tilde{O}_{xzz}^{\text{(m)}}(z_1) + \cdots \Bigr] = \widehat{\Delta E}(0)\cos(k z_1) + j\widehat{\Delta H}(0)\sin(k z_1), \\[2ex]
    \frac{\eta}{2}&\Bigl[\tilde{P}_x(z_2) +  \tilde{Q}_{yz}^{\text{(m)}}(z_2) +\tilde{O}_{yzz}^{\text{(e)}}(z_2) + \cdots \Bigr] = \widehat{\Delta H}(0)\cos(k z_2) + j\widehat{\Delta E}(0)\sin(k z_2),
\end{align}
\end{subequations}
where we have chosen the different positions $z_1$ and $z_2$ for these two equations since they may be separated according to their respective parities.

Now, by comparing both sides of the equations in~\eqref{eq:conventional_expansion}, we can pinpoint the exact spatial origins where the dipole moments alone are enough to satisfy the response, effectively minimizing the need for higher-order multipoles. This translates mathematically into seeking the optimal origins $z_1^{\text{opt}}$ and $z_2^\text{opt}$ that globally minimize the non-dipolar multipole moments. To do so, we bring the dipolar contributions present on the left-hand sides of~\eqref{eq:conventional_expansion} to the corresponding right-hand sides of these equations. What remains on left-hand sides of these equations is therefore only the higher-order multipole moments to be minimized. The corresponding minimization functions are
\begin{subequations}\label{eq:optimal_z}
\begin{align}
    \left| \widehat{\Delta E}(0)\cos(k z_1^{\text{opt}}) + j\widehat{\Delta H}(0)\sin(k z_1^{\text{opt}}) - \frac{1}{2}\tilde{M}_y(z_1^{\text{opt}}) \right| &= 0, \\[1ex]
    \left| \widehat{\Delta H}(0)\cos(k z_2^{\text{opt}}) + j\widehat{\Delta E}(0)\sin(k z_2^{\text{opt}}) - \frac{\eta}{2}\tilde{P}_x(z_2^{\text{opt}}) \right| &= 0.
\end{align}
\end{subequations}

Importantly, inspecting~\eqref{eq:optimal_z} reveals that it is not necessary to perform a \emph{full} multipolar decomposition at every position $z$. Instead, the complete decomposition only needs to be carried out once at the reference point, which is at the CM. Subsequently, only the \emph{dipolar} moments need to be evaluated for different values of $z_1$ and $z_2$ in order to minimize~\eqref{eq:optimal_z} and determine the optimal positions $z_1^{\text{opt}}$ and $z_2^{\text{opt}}$.

\begin{figure}[H]
    \centering
    \includegraphics[width=\textwidth]{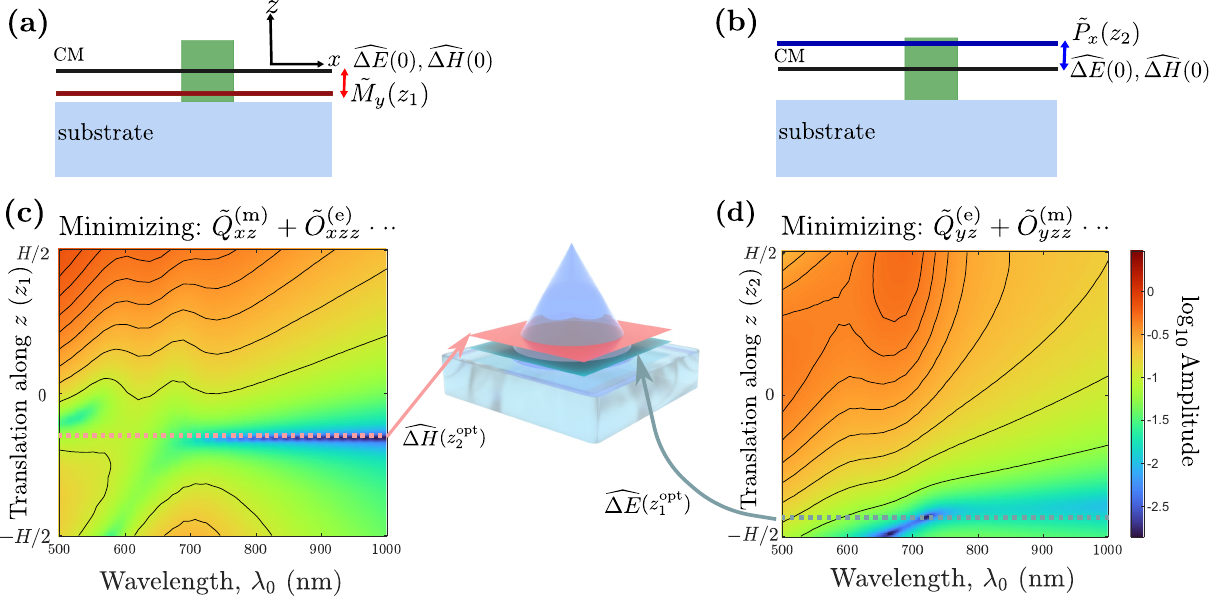}
    \caption{Finding the optimal origins for both parities of multipoles. In (a) and (b), all multipole moments are initially computed at the CM, which yields $\widehat{\Delta E}(0)$ and $\widehat{\Delta H}(0)$. Then, to find the positions $z_1^\text{opt}$ and $z_2^\text{opt}$ that minimize the higher-order multipole moments, the electric and magnetic dipole moments are separately computed at different positions along the $z$-axis. (c) and (d) plot the left-hand sides of~\eqref{eq:optimal_z} versus wavelength and $z$-positions. The optimal origins are highlighted by dashed lines.}
    \label{fig_1:design}
\end{figure}

To illustrate this methodology, we now consider the substrated cone with $R_\text{top}=0$ shown in Fig.~\ref{fig_2:design} and use~\eqref{eq:optimal_z} to minimize all of its higher-order multipoles. Figure~\ref{fig_1:design}a and~\ref{fig_1:design}b illustrate the structural configuration, where $\widehat{\Delta E}(0)$ and $\widehat{\Delta H}(0)$ are computed at the CM and the dipole moments are computed at arbitrary $z$-positions. The resulting left-hand sides of~\eqref{eq:optimal_z} are plotted in Fig.~\ref{fig_1:design}c and~\ref{fig_1:design}d.
In these plots, we clearly see that the higher-order multipole moments are globally minimized at different $z$-positions depending on the parity of the multipoles. This confirms that the concept of multipoles parity-splitting introduced in~\cite{kildishev2025art} can indeed be generalized to the case of metasurfaces. In this example, both optimal origins are below the CM at $z_1^{\text{opt}}=-56~\text{nm}$ and $z_2^{\text{opt}}=-112~\text{nm}$.

\begin{figure}[H]
    \centering
    \includegraphics[width=0.8\textwidth]{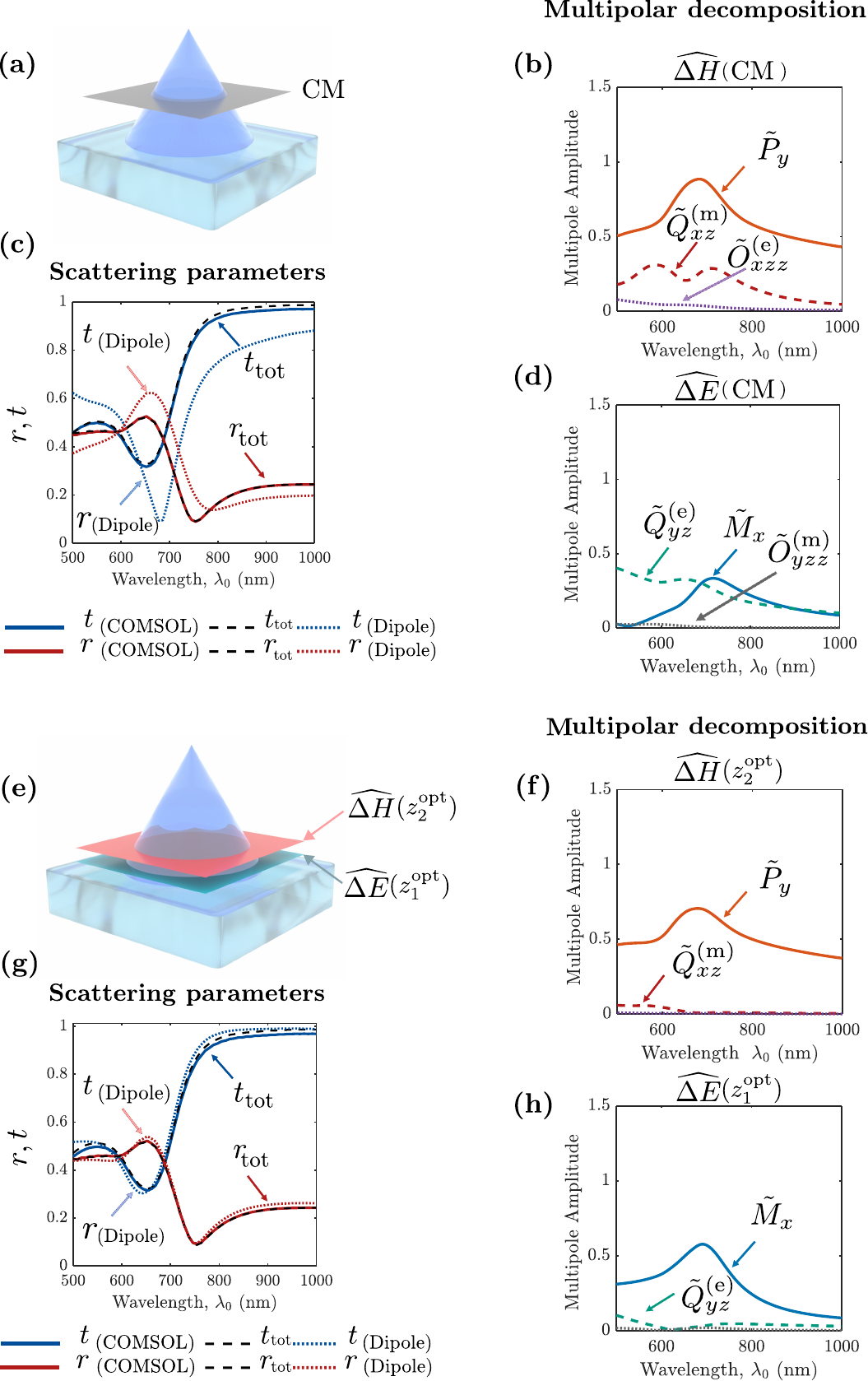}
    \caption{Modeling of the scattering response of an amorphous silicon cone ($R_\text{top}=0$) on top of a glass substrate. In (a)-(d), the modeling is performed with both GSTCs relations placed at the CM. (b) and (d) plot the corresponding multipolar decomposition. In (e)-(h), the modeling is performed with parity split GSTCs placed at the optimal positions that minimize higher-order multipoles. (f) and (h) plot the corresponding multipolar decomposition. (c) and (g) Comparison of the COMSOL simulated scattering parameters with the ones retrieved using~\eqref{eq:RT_final_substrate} with all multipoles moments (dashed lines) and with only dipole moments (dotted lines).}
    \label{fig_4:design}
\end{figure}

Let us now compare the multipolar GSTCs modeling technique when both origins are located at the CM with the case where they are placed at the optimal positions $z_1^{\text{opt}}$ and $z_2^{\text{opt}}$. We begin with the configuration in which the GSTCs are defined at the CM, as illustrated in Fig.~\ref{fig_4:design}a. The corresponding multipolar decompositions are presented in Fig.~\ref{fig_4:design}b and Fig.~\ref{fig_4:design}d for both parities of multipoles. As observed, performing the multipolar decomposition at the CM leads to the presence of significant quadrupolar and even octupolar contributions. Then, in Fig.~\ref{fig_4:design}c, we compare the scattering parameters obtained from COMSOL simulations (solid lines) with those retrieved using our GSTCs formalism when including either all multipole moments (dashed lines) or only the dipolar contributions (dotted lines). Because the higher-order multipoles contribute substantially to the response, they cannot be neglected. Consequently, only the scattering parameters reconstructed using the full set of multipole moments remain in good agreement with the COMSOL simulations, whereas the dipole-only reconstruction fails to reproduce the expected scattering response.

Now, we consider the optimal origin points $z_1^{\text{opt}}=?$ and $z_2^{\text{opt}}=?$ that were found previously, and represent the corresponding GSTCs in Fig.~\ref{fig_4:design}e. The corresponding multipolar decomposition are shown in Fig.~\ref{fig_4:design}f and~\ref{fig_4:design}h for both parities of multipoles. This time, we see that the octupolar responses are essentially negligible, while the quadrupolar ones have been significantly reduced in amplitude compared to the dipolar ones. In Fig.~\ref{fig_4:design}g, we compare the COMSOL simulated scattering parameters (solid lines) with the ones obtained using our GSTCs technique that include all multipole moments (dashed line) or only dipole moments (dotted lines). As can be seen, both retrieved scattering parameters are in excellent agreement with the COMSOL simulations. This demonstrates the capability of our approach to accurately model a complex electromagnetic structure using only electric and magnetic dipole moments.

\newpage

\subsection{Multipolar Minimization via In-Plane Origin Translation}
\label{sec_inplane}

So far, we have only considered cases where the origin translation was along the $z$-direction. Now, we shall discuss cases where the origin translation is within the $xy$-plane. In this case, because we are not translating along the $z$-direction and that the incident wave is normally impinging on the metasurface, there is no in-plane phase variation to take into account as was needed to define the scattering parameters in~\eqref{eq:ms_intrinsic}. This means that the GSTCs, and the associated scattering parameters, remain invariant under such in-plane translations. In this case, equations~\eqref{eq:optimal_z} can be generalized by considering the vector position $\mathbf{r}_{\text{opt}}$, as
\begin{subequations}\label{eq:optimal_zx}
\begin{align}
    \left| \widehat{\Delta E}(\mathbf{0})\cos(k_{z} z_1^{\text{opt}})+ j\widehat{\Delta H}(\mathbf{0})\sin(k_{z} z_1^{\text{opt}})- \frac{1}{2}\tilde{M}_x(\mathbf{r}_n^\text{opt}) \right| &= 0, \\[1ex]
    \left| \widehat{\Delta H}(\mathbf{0})\cos(k_{z} z_2^{\text{opt}})+ j\widehat{\Delta E}(\mathbf{0})\sin(k_{z} z_2^{\text{opt}})- \frac{\eta}{2}\tilde{P}_y(\mathbf{r}_n^\text{opt}) \right| &= 0,
\end{align}
\end{subequations}
where $\mathbf{0} = (0,0,0)$ denotes the position of the reference point (in our case the CM) and $\mathbf{r}_n^\text{opt}=(x_n^{\text{opt}},y_n^{\text{opt}},z_n^{\text{opt}})$ with $n=\{1,2\}$.

If the coordinate origin is defined at the center of the object, which is usually the case for $z$-symmetric structures, such that we do not need to translate along the $z$-axis ($z=0$), the argument of the trigonometric functions becomes zero. Under this condition, relations~\eqref{eq:optimal_zx} reduce to
\begin{subequations}\label{eq:optimal_normal}
\begin{align}
    \left| \widehat{\Delta E}(0) - \frac{1}{2}\tilde{M}_x(x_1^{\text{opt}},y_1^{\text{opt}}) \right| &= 0, \\[1ex]
    \left| \widehat{\Delta H}(0) - \frac{\eta}{2}\tilde{P}_y(x_2^{\text{opt}},y_2^{\text{opt}}) \right| &= 0.
\end{align}
\end{subequations}
This formula implies that, in order to minimize higher-order multipoles, we need to compute all multipole moments at the CM and then find the minima of~\eqref{eq:optimal_normal} by computing only the dipolar contributions at various $x$ and $y$ positions.



In the following, we study a structure that is asymmetry within the $xy$-plane and exhibits a double quasi-BIC resonance, where each resonance is dominated by a specific multipolar parity, though some parity mixing inevitably occurs due to the broken symmetry of the structure. Because the structure remains symmetric along the $z$-axis, the multipole moments do not vary significantly in this direction. Moreover, BICs are typically associated with planar modes that share this $z$-axis symmetry. Therefore, we only consider spatial translations along the $x$- and $y$-directions to obtain the optimal origin points. As before, our goal is to find the optimal in-plane points that minimize the contributions from higher-order multipole moments.

\begin{figure}[H]
    \centering
    \includegraphics[width=\textwidth]{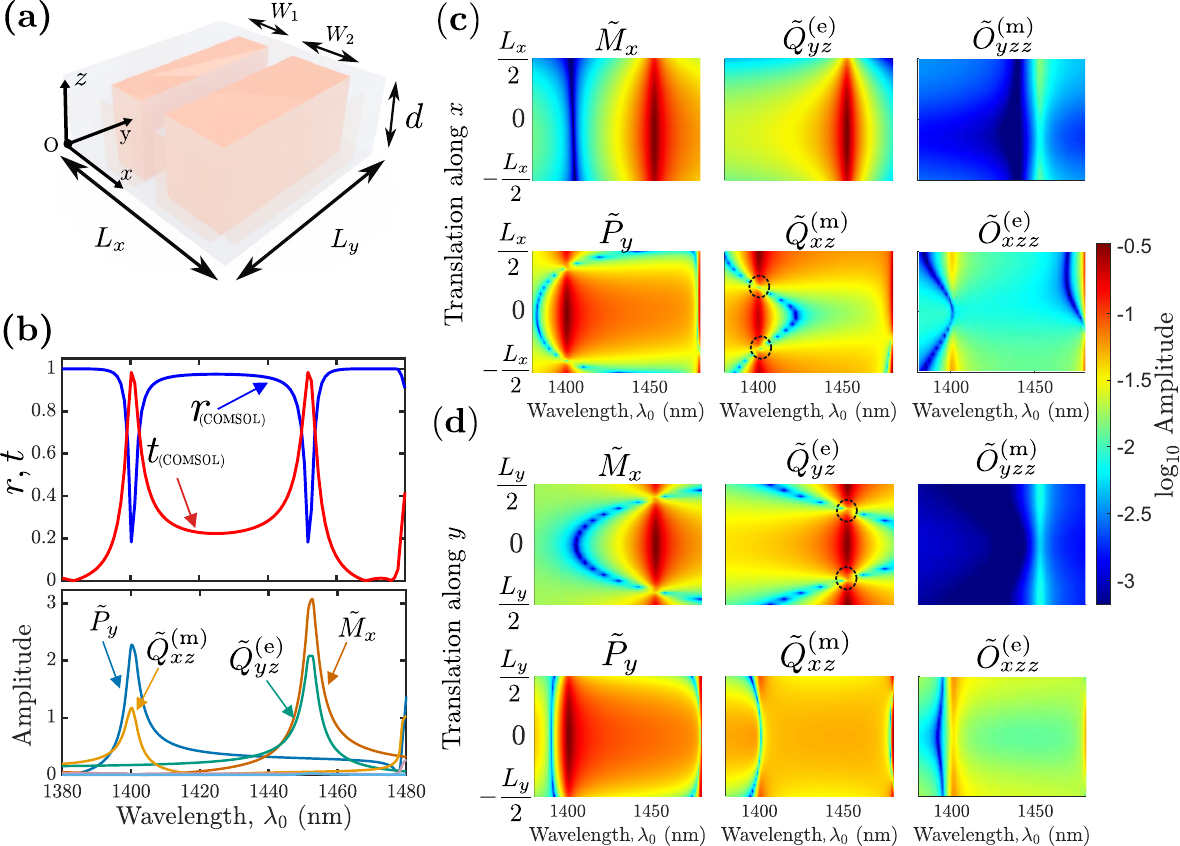}
    \caption{Multipolar modeling of an asymmetric metasurface exhibiting double quasi-BIC resonances. (a)~Unit cell geometry composed of two asymmetric Ti$\text{O}_2$ blocks embedded in Si$\text{O}_2$ with the corresponding geometry parameters as $L_x=L_y=900~\text{nm},~~d=320~\text{nm},~~W_1=200~\text{nm}, ~~W_2=350~\text{nm}$. (b)~COMSOL simulated reflectance and transmittance (top) and corresponding multipolar decomposition (bottom) with the system of coordinates, $\mathbf{O}$, placed in the middle of the unit cell, i.e., $\mathbf{O}=[L_x/2,L_y/2,d/2]$. The amplitudes of the multipoles, for various in-plane translations of the system of coordinates, are plotted in (c) for $\mathbf{O}=[x,L_y/2,d/2]$ and in (d) for $\mathbf{O}=[2L_x/10,y,d/2]$. The black dashed circles indicate nodes of the quadrupolar responses where the optimal positions are selected along $x$ and $y$.}
    \label{fig_6:design}
\end{figure}

We investigate a unit cell composed of two asymmetric dielectric blocks, as depicted in Fig.~\ref{fig_6:design}a. The different width of the blocks leads to a double resonance response. In Fig.~\ref{fig_6:design}b, we plot the COMSOL simulated scattering parameters and the corresponding multipole moments computed at the center of the unit cell (CU). As can be seen, the scattering response in Fig.~\ref{fig_6:design}b shows the presence of two resonances. By inspecting the multipolar decomposition, it is evident that each resonance is dominated by a different parity of multipoles. Although a slight mixing of the parities is detectable due to the broken symmetry. We also observe a strong contribution from quadrupolar terms within each resonance with negligible octupolar contribution. In Fig.~\ref{fig_6:design}c and~\ref{fig_6:design}d, we plot the corresponding multipole moments computed at $z=d/2$ but for different positions along $x$ and $y$.
Interestingly, we see in Fig.~\ref{fig_6:design}c that the dipole and quadrupole moments, $\tilde{M}_x$ and $\tilde{Q}_{yz}^\text{(e)}$, exhibit distinct zero-crossings (nodes) when the origin is translated along the $y$-direction. By duality, we see in Fig.~\ref{fig_6:design}d that the same phenomenon occurs for the dipole and quadrupole moments, $\tilde{P}_y$ and $\tilde{Q}_{xz}^\text{(m)}$ when the origin is translated along the $x$-direction. This will allow us to select the positions where the quadrupolar responses are minimized. Note that the octupolar responses are essentially negligible in this structure and we therefore do not discuss them any further. 

\begin{figure}[H]
    \centering
    \includegraphics[width=\textwidth]{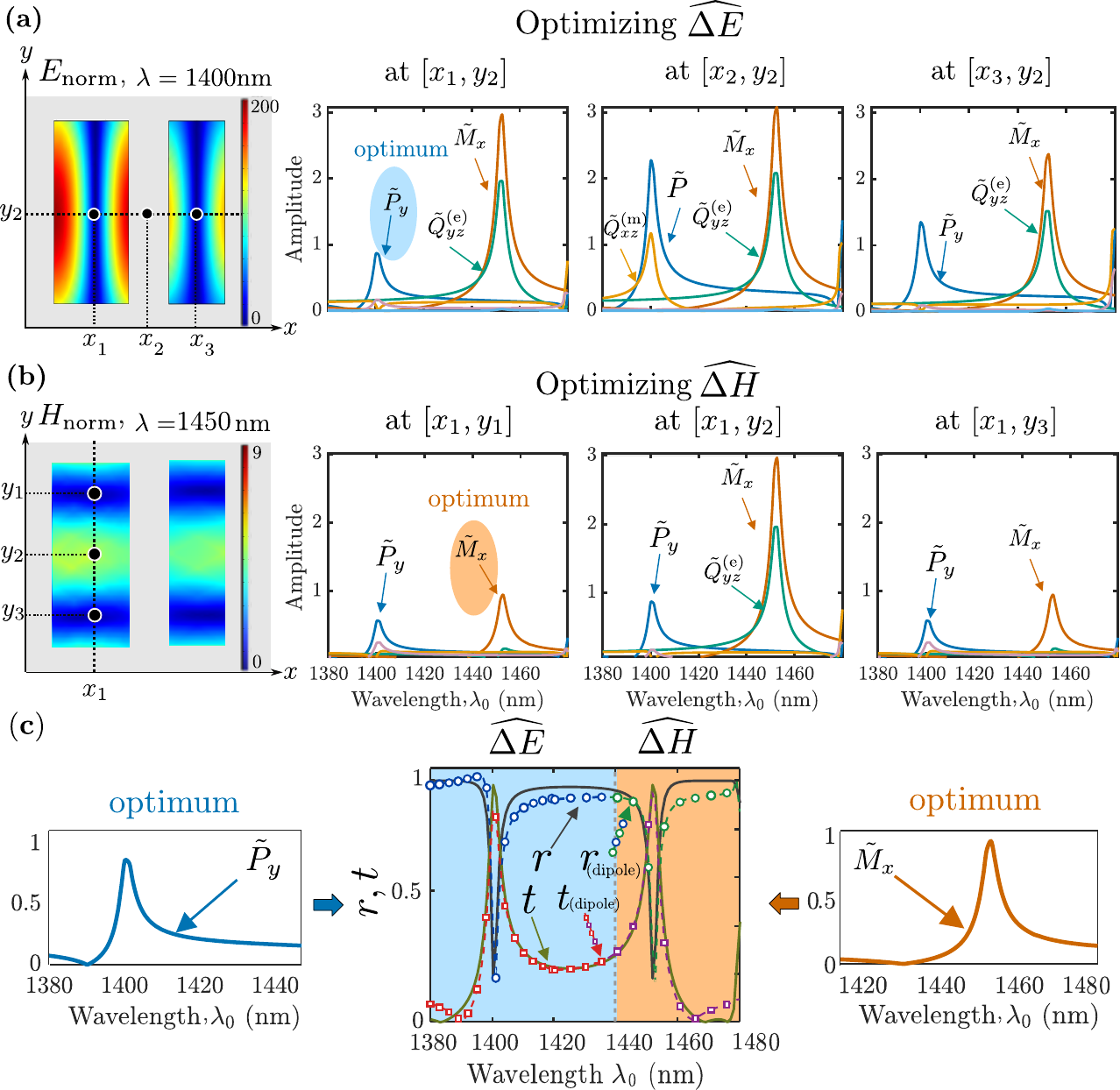}
    \caption{Multipolar minimization via in-plane origin translation for the symmetry-broken double-block metasurface. (a) Normalized electric field distribution at the first quasi-BIC resonance ($\lambda = 1400$~nm) and corresponding multipole moments evaluated at different $x$-positions along a fixed $y=L_y/2$. The $x$ coordinates are $[x_1,x_2,x_3]=[2L_x/10,L_x/2,7L_x/10]$. Translating the coordinate origin to $[x_1, y_2]$ optimizes the even parity multipole moments to only electric dipole $\tilde{P}_y$. (b) Normalized magnetic field distribution at the second quasi-BIC resonance ($\lambda = 1450$~nm) and corresponding multipole moments evaluated at different $y$-positions along the fixed optimal $x_1$ position with $[y_1,y_2,y_3]=[2L_y/10,L_y/2,8L_y/10]$. Translating the origin to the nodal position $[x_1, y_1]$ optimizes the odd parity multipole moments to only magnetic dipole $\tilde{M}_x$. (c) Retrieval of the scattering parameters using only the extracted optimal dipolar moments. By evaluating the boundary conditions independently at their respective parity-split optimum positions for each resonance, the analytically reconstructed reflection and transmission (circles) show excellent agreement with the full-wave numerical simulations (solid lines) around each resonance.}
    \label{fig_7:design}
\end{figure}

We now examine the multipole moments at each of the identified nodes as well as at the CU for the two quasi-BIC resonances in Fig.~\ref{fig_6:design}b, which occur at $\lambda=1400$~nm and $\lambda=1450$~nm, respectively. For the first quasi-BIC resonance, we plot the norm of the electric field inside the dielectric blocks in Fig.~\ref{fig_7:design}a along with the corresponding multipolar decomposition at the two nodes highlighted in Fig.~\ref{fig_6:design}c and at the CU. For the multipolar decomposition at the CU, we see that both dipolar and quadrupolar responses are needed to model both resonances. However, the quadrupolar response $\tilde{Q}_{xz}^\text{(m)}$ is minimized at both nodes, which interestingly coincides with the electric field nodal lines. Furthermore, due to the structural asymmetry, the multipolar responses differ between the two nodes. Specifically, the contribution from the dipolar response $\tilde{P}_{y}$ is optimized only at the first node, located at $(x_1,L_y/2,d/2)$. This specific point therefore serves as the optimal origin for the even parity component of the GSTCs. 

For the second quasi-BIC resonance, we plot the norm of the magnetic  field inside the dielectric blocks in Fig.~\ref{fig_7:design}b along with the corresponding multipolar decomposition at the two nodes highlighted in Fig.~\ref{fig_6:design}d and the new reference center, which corresponds to the previously identified optimal point $(x_1,L_y/2,d/2)$ to avoid the appearance of higher-order even multipoles due to parity mixing. Because of electromagnetic duality, we expect to observe a minimum for $Q_{yz}^{(\text{e})}$ under translation along $y$. Therefore, we translate the origin along the $y$-direction while keeping the $x$-coordinate fixed at $x_1$. Because this translation occurs along the symmetry line, two distinct minima emerge for $Q_{yz}^{(\text{e})}$, which interestingly match with the nodal lines of the magnetic field distribution. Therefore, both positions,~$(x_1,y_3,d/2)$ and~$(x_3,y_3,d/2)$, are valid candidates to serve as the optimal origin for the magnetic dipole. In the following, we will only consider the first position.

We now use these two optimal origins, located at $(x_1,L_y/2,d/2)$ and $(x_1,y_3,d/2)$, to retrieve the overall metasurface scattering parameters. To do so, we directly use relations~\eqref{eq:ms_intrinsic}, since the metasurface is fully embedded in Si$\text{O}_2$ (no substrate), and plot the resulting scattering parameters in Fig.~\ref{fig_7:design}c. Note that instead of simultaneously using both electric and magnetic dipoles, and the corresponding even and odd parts of the GSTCs, to model this double resonance response, we rather model each resonance separately using the GSTC relation with the appropriate parity. Specifically, we model the first resonance only using the electric dipole response and we transition away from it as its contribution approaches zero. Then, for the second resonance, we switch to the magnetic dipole response as its amplitude begins to increase. This transition occurs at a wavelength of $\lambda = 1440$~nm. We do this because we know that this double quasi-BIC resonance is split into even and odd parities, as previously explained.

As can be seen, at both resonance frequencies, the scattering parameters of the metasurface are accurately modeled using only the electric and magnetic dipoles. Away from the resonance, the mixing of spatial parities becomes more prominent. Consequently, achieving high accuracy in these off-resonance regions would require evaluating both sets of dipolar moments simultaneously.


\section{Conclusion}

In this work, we present a method to simplify the multipolar modeling of metasurfaces without losing precision. By taking advantage of the fact that multipole moments depend on the choice of the coordinates origin, we show how to find optimal positions that minimize the contribution of multipole moments beyond the dipolar ones. 

The main idea is based on separating the metasurface response into even and odd parity field discontinuities, which can then be evaluated at different physical positions. Because the optimal origins for minimizing high-order multipoles are generally different for each parity, evaluating both GSTC relations at the same origin forces the model to include a large number of high-order multipoles. By splitting the GSTCs and relating them to these distinct optimal origins, we can accurately reconstruct the metasurface scattering response using only dipole moments. This approach provides an efficient and physically clear framework that reduces modeling complexity, making it a useful tool for the design and analysis of complex metasurfaces.

We validate our analytical approach using two different examples. First, we model a series of dielectric cones with different levels of conicity on a substrate. For the cone with $R_{top}=0$, we show that a combination of electric and magnetic dipole moments are sufficient to fully retrieve the scattering parameters of the metasurface. 

In the second example, we apply our method to a symmetry-broken double-block metasurface supporting a double quasi-BIC resonance, where we use in-plane translations along the $x$- and $y$-directions to find the optimal origins. Since each quasi-BIC resonance corresponds to a specific field parity, we are able to show that a single GSTC relation, and thus a single dipole moment with the right parity, is sufficient to model that resonance. By stitching together the scattering response of both resonances, we are finally able to fully reconstruct the overall metasurface response.

For future work, this method could be expanded to include oblique incidence. When a wave is incident on the metasurface at an angle, it introduces a phase gradient along the transverse direction and excites normal multipolar components. Because multipole moments change with the illumination angle, the optimal origin positions for the even and odd GSTCs will become angle-dependent. Adjusting our translation rules to account for this variation is an important future challenge. Furthermore, reducing the metasurface response to only dipole moments is highly beneficial for T-matrix formulations and inverse design approaches, where avoiding higher-order multipoles simplifies the computations. Ultimately, these extensions will make this method a complete, purely dipolar framework capable of handling complex metasurface designs in any general environment.

\bibliographystyle{ieeetr}
\bibliography{references}

\clearpage 


\renewcommand{\theequation}{S\arabic{equation}}
\renewcommand{\thefigure}{S\arabic{figure}}
\renewcommand{\thesection}{S\arabic{section}}
\renewcommand{\thetable}{S\arabic{table}}

\begin{center}
    \Large \textbf{Supplementary Material} \\
    \vspace{1ex}
    \large \textbf{}
\end{center}
\vspace{2em}

\section{Multipolar supplementary materials}\label{sec:Multipole_supp}

\begin{table}[H]
    \centering
    \caption{Multipole representations for electric and magnetic multipoles.}
    \label{tab:multipoles} 
    \small 
    \renewcommand{\arraystretch}{3}
    \begin{tabular*}{0.916\linewidth}{|c|l|} 
        \hline \hline
        Order & Electric Multipoles Representation \\
        \hline
        1 & $\displaystyle P_i = -\frac{1}{i\omega A} \int_V \, j_0(kr)J_i \ dV + \frac{k^2}{2} \int_V  \frac{j_2(kr)}{(kr)^2} \left[ 3(\mathbf{r} \cdot \mathbf{J})r_i - r^2 J_i \right] dV$ \\
        \hline
        2 & $\begin{aligned}
             Q_{ij}^\text{(e)} = -\frac{3}{i\omega A} \Bigg[& \int_V \frac{j_1(kr)}{kr}\left[ 3(r_j J_i + r_i J_j) - 2(\mathbf{r} \cdot \mathbf{J})\delta_{ij} \right] dV \\
             & + 2k^2 \int_V \frac{j_3(kr)}{(kr)^3} \left[ 5r_i r_j (\mathbf{r} \cdot \mathbf{J}) - (r_i J_j + r_j J_i)r^2 - r^2(\mathbf{r} \cdot \mathbf{J})\delta_{ij} \right] dV \Bigg]
           \end{aligned}$ \\
        \hline
        3 & $\displaystyle O_{ijk}^\text{(e)} = -\frac{15}{i\omega A} \int_V \frac{j_2(kr)}{(kr)^2} \left[ J_i r_j r_k + r_i J_j r_k + r_i r_j J_k - \left(\delta_{ij} V_k^\text{(e)} + \delta_{ik} V_j^\text{(e)} + \delta_{jk} V_i^\text{(e)}\right) \right] dV$ \\
        \hline 
        4 & $\displaystyle X_{ijkl}^\text{(e)} = -\frac{105}{i\omega A} \int_V \frac{j_4(kr)}{(kr)^4} \left[ J_i r_j r_k r_l + r_i J_j r_k r_l + r_i r_j J_k r_l + r_i r_j r_k J_l - B_{ijkl}^\text{(e)} \right] dV$ \\
        
        \hline \hline
        Order & Magnetic Multipoles Representation \\
        \hline
        1 & $\displaystyle M_i = \frac{3}{2A} \int_V \frac{j_1(kr)}{kr}\, (\mathbf{r} \times \mathbf{J})_i \ dV$ \\
        \hline
        2 & $\displaystyle Q_{ij}^\text{(m)} = \frac{15}{A} \int_V   \frac{j_2(kr)}{(kr)^2}\left[ r_i(\mathbf{r} \times \mathbf{J})_j + r_j(\mathbf{r} \times \mathbf{J})_i \right] dV $\\
        \hline
       3 & $\begin{aligned}
             O_{ijk}^\text{(m)} = \frac{105}{4A} \int_V \frac{j_3(kr)}{(kr)^3} \Big[ & \epsilon_{ilm}r_l J_m r_j r_k + r_i \epsilon_{jlm}r_l J_m r_k + r_i r_j \epsilon_{klm}r_l J_m \\
             & - \left(\delta_{ij} V_k^\text{(m)} + \delta_{ik} V_j^\text{(m)} + \delta_{jk} V_i^\text{(m)}\right) \Big] dV
           \end{aligned}$ \\
       \hline
       4 & $\begin{aligned}
             X_{ijkl}^\text{(m)} = \frac{189}{A} \int_V \frac{j_4(kr)}{(kr)^4} \Big[ & \epsilon_{ipq}r_p J_q r_j r_k r_l + r_i \epsilon_{jpq}r_p J_q r_k r_l \\
             & + r_i r_j \epsilon_{kpq}r_p J_q r_l + r_i r_j r_k \epsilon_{lpq}r_p J_q - B_{ijkl}^\text{(m)} \Big] dV
           \end{aligned}$ \\
       \hline \hline
    \end{tabular*}
\end{table}

\begin{table}[htbp]
    \centering
    \renewcommand{\arraystretch}{2} 
    \begin{tabular}{l | l}
        \hline
        ET for electric & $\begin{aligned}
            B_{ijkl}^{(e)} &= \frac{1}{7} \left( \delta_{ij}\hat{C}_{kl}^{(e)} + \delta_{ik}\hat{C}_{jl}^{(e)} + \delta_{il}\hat{C}_{jk}^{(e)} + \delta_{jk}\hat{C}_{il}^{(e)} + \delta_{jl}\hat{C}_{ik}^{(e)} + \delta_{kl}\hat{C}_{ij}^{(e)} \right), \\
            \hat{\mathbf{C}}^{(e)} &= 2(\mathbf{J} \cdot \mathbf{r})\mathbf{r}\mathbf{J}\mathbf{r} + r^2(\mathbf{J}\mathbf{r} + \mathbf{r}\mathbf{J}) - \frac{4}{3}r^2(\mathbf{r} \cdot \mathbf{J})\bar{\bar{I}}, \quad \mathbf{V}^{(e)} = \frac{1}{5}\left[ 2(\mathbf{r} \cdot \mathbf{J})\mathbf{r} + r^2\mathbf{J} \right]
        \end{aligned}$ \\
        \hline
        ET for magnetic & $\begin{aligned}
            B_{ijkl}^{(m)} &= \frac{1}{7A} \left( \delta_{ij}\hat{C}_{kl}^{(m)} + \delta_{ik}\hat{C}_{jl}^{(m)} + \delta_{il}\hat{C}_{jk}^{(m)} + \delta_{jk}\hat{C}_{il}^{(m)} + \delta_{jl}\hat{C}_{ik}^{(m)} + \delta_{kl}\hat{C}_{ij}^{(m)} \right), \\
            \hat{\mathbf{C}}^{(m)} &= r^2\left[ (\mathbf{r} \times \mathbf{J})\mathbf{r} + \mathbf{r}(\mathbf{r} \times \mathbf{J}) \right], \quad \mathbf{V}^{(m)} = \frac{1}{5}r^2(\mathbf{r} \times \mathbf{J})
        \end{aligned}$ \\
        \hline
    \end{tabular}
    \vspace{0.2cm}
    \caption{Electric and magnetic extra tensor terms used in the expansion.}
    \label{tab:extra_tensors}
\end{table}

\begin{table}[H]
    \centering
    \renewcommand{\arraystretch}{2.2} 
    \caption{TE Normalized Multipoles coefficients categorized by Parity}
    \vspace{1ex}
    \begin{tabular}{|c|c|c|c|}
        \hline
        \textbf{Original Multipole} & \textbf{Tilde Form} & \textbf{Original Multipole} & \textbf{Tilde Form} \\
        \hline
        \multicolumn{4}{|c|}{\textbf{Even Parity ($\bm{\Delta E_y}^{\text{TE}}$)}} \\
        \hline
        $i\omega\mu M_y$ & $\tilde{M}_y$ & $\frac{k^2}{2\epsilon} Q_{xz}^{(\text{e})}$ & $\tilde{Q}_{xz}^{(\text{e})}$ \\
        \hline
        $-\frac{\omega\mu}{2} k_x Q_{yy}^{(\text{e})}$ & $\tilde{Q}_{yy}^{(\text{e})}$ & $-\frac{\omega\mu}{2} k_x Q_{zz}^{(\text{e})}$ & $\tilde{Q}_{zz}^{(\text{e})}$ \\
        \hline
        $\frac{i k^2 k_x}{3\epsilon} O_{yzx}^{(\text{e})}$ & $\tilde{O}_{yzx}^{(\text{e})}$ & $\frac{i\omega k_z^2}{6} O_{yzz}^{(\text{m})}$ & $\tilde{O}_{yzz}^{(\text{m})}$ \\
        \hline
        \multicolumn{4}{|c|}{\textbf{Odd Parity ($\bm{\Delta H_x}^{\text{TE}}$)}} \\
        \hline
        $-i\omega P_x$ & $\tilde{P}_x$ & $-i k_x M_z$ & $\tilde{M}_z$ \\
        \hline
        $\frac{k_x^2 - k_z^2}{2} Q_{yz}^{(\text{m})}$ & $\tilde{Q}_{yz}^{(\text{m})}$ & $-\frac{\omega k_x}{2} Q_{yx}^{(\text{m})}$ & $\tilde{Q}_{yx}^{(\text{m})}$ \\
        \hline
        $-\frac{i\omega k_x^2}{6} O_{xyy}^{(\text{e})}$ & $\tilde{O}_{xyy}^{(\text{e})}$ & $\frac{i\omega k_z^2}{6} O_{xzz}^{(\text{e})}$ & $\tilde{O}_{xzz}^{(\text{e})}$ \\
        \hline
    \end{tabular}
    \label{tab:te_multipoles}
\end{table}

\begin{equation} 
\label{eq_deltaE_parallel}
\begin{aligned}
\Delta \mathbf{E}_{\parallel} &= \frac{-1}{k_z \omega \epsilon} \bigg\{ (-\mathbf{k}_{\parallel}\mathbf{k}_{\parallel} + \partial_z \partial_z + k^2 \mathbf{I}) \cdot \Big[ \partial_z \times \mathbf{m} + \frac{i\omega}{2}\partial_z \cdot \mathbf{Q}^{(e)} \\
&\quad - \frac{1}{2}i\mathbf{k}_{\parallel} \times (\partial_z \cdot \mathbf{Q}^{(m)}) - \frac{1}{2}\partial_z \times (i\mathbf{k}_{\parallel} \cdot \mathbf{Q}^{(m)}) + \frac{i\omega}{6}i\mathbf{k}_{\parallel} \cdot (\partial_z \cdot \mathbf{O}^{(e)}) \\
&\quad + \frac{i\omega}{6}\partial_z \cdot (i\mathbf{k}_{\parallel} \cdot \mathbf{O}^{(e)}) + \frac{1}{2}i\mathbf{k}_{\parallel} \times (i\mathbf{k}_{\parallel} \cdot \partial_z \cdot \mathbf{O}^{(m)}) \\
&\quad + \frac{1}{2}i\mathbf{k}_{\parallel} \times (\partial_z \cdot i\mathbf{k}_{\parallel} \cdot \mathbf{O}^{(m)}) + \frac{1}{2}\partial_z \times (i\mathbf{k}_{\parallel} \cdot i\mathbf{k}_{\parallel} \cdot \mathbf{O}^{(m)}) \\
&\quad + \frac{1}{2}\partial_z \times (\partial_z \cdot \partial_z \cdot \mathbf{O}^{(m)}) \Big] \\
&\quad + (i\mathbf{k}_{\parallel}\partial_z + \partial_z i\mathbf{k}_{\parallel}) \cdot \Big[ -i\omega\mathbf{p} + i\mathbf{k}_{\parallel} \times \mathbf{m} + \frac{i\omega}{2}i\mathbf{k}_{\parallel} \cdot \mathbf{Q}^{(e)} \\
&\quad - \frac{1}{2}i\mathbf{k}_{\parallel} \times (i\mathbf{k}_{\parallel} \cdot \mathbf{Q}^{(m)}) - \frac{1}{2}\partial_z \times (\partial_z \cdot \mathbf{Q}^{(m)}) - \frac{i\omega}{6}i\mathbf{k}_{\parallel} \cdot (i\mathbf{k}_{\parallel} \cdot \mathbf{O}^{(e)}) \\
&\quad - \frac{i\omega}{6}\partial_z \cdot (\partial_z \cdot \mathbf{O}^{(e)}) + \frac{1}{2}i\mathbf{k}_{\parallel} \times (i\mathbf{k}_{\parallel} \cdot i\mathbf{k}_{\parallel} \cdot \mathbf{O}^{(m)}) \\
&\quad + \frac{1}{2}i\mathbf{k}_{\parallel} \times (\partial_z \cdot \partial_z \cdot \mathbf{O}^{(m)}) + \frac{1}{2}\partial_z \times (i\mathbf{k}_{\parallel} \cdot \partial_z \cdot \mathbf{O}^{(m)}) \\
&\quad + \frac{1}{2}\partial_z \times (\partial_z \cdot i\mathbf{k}_{\parallel} \cdot \mathbf{O}^{(m)}) \Big] \bigg\}.
\end{aligned}
\end{equation}

\begin{equation} \label{eq:deltaH_parallel}
\begin{aligned}
\Delta \mathbf{H}_{\parallel} &= \frac{-1}{ik_z} \bigg\{ i\mathbf{k}_{\parallel} \times \Big[ \partial_z \times \mathbf{m} + \frac{i\omega}{2}\partial_z \cdot \mathbf{Q}^{(e)} - \frac{1}{2}i\mathbf{k}_{\parallel} \times (\partial_z \cdot \mathbf{Q}^{(m)}) \\
&\quad - \frac{1}{2}\partial_z \times (i\mathbf{k}_{\parallel} \cdot \mathbf{Q}^{(m)}) - \frac{i\omega}{6}i\mathbf{k}_{\parallel} \cdot (\partial_z \cdot \mathbf{O}^{(e)}) - \frac{i\omega}{6}\partial_z \cdot (i\mathbf{k}_{\parallel} \cdot \mathbf{O}^{(e)}) \\
&\quad + \frac{1}{2}i\mathbf{k}_{\parallel} \times (i\mathbf{k}_{\parallel} \cdot \partial_z \cdot \mathbf{O}^{(m)}) + \frac{1}{2}i\mathbf{k}_{\parallel} \times (\partial_z \cdot i\mathbf{k}_{\parallel} \cdot \mathbf{O}^{(m)}) \\
&\quad + \frac{1}{2}\partial_z \times (i\mathbf{k}_{\parallel} \cdot i\mathbf{k}_{\parallel} \cdot \mathbf{O}^{(m)}) + \frac{1}{2}\partial_z \times (\partial_z \cdot \partial_z \cdot \mathbf{O}^{(m)}) \Big] \\
&\quad + \partial_z \times \Big[ -i\omega\mathbf{p} + i\mathbf{k}_{\parallel} \times \mathbf{m} + \frac{i\omega}{2}i\mathbf{k}_{\parallel} \cdot \mathbf{Q}^{(e)} + \frac{1}{2}\mathbf{k}_{\parallel} \times (\mathbf{k}_{\parallel} \cdot \mathbf{Q}^{(m)}) \\
&\quad - \frac{1}{2}\partial_z \times (\partial_z \cdot \mathbf{Q}^{(m)}) + \frac{i\omega}{6}\mathbf{k}_{\parallel} \cdot (\mathbf{k}_{\parallel} \cdot \mathbf{O}^{(e)}) - \frac{i\omega}{6}\partial_z \cdot (\partial_z \cdot \mathbf{O}^{(e)}) \\
&\quad + \frac{1}{2}i\mathbf{k}_{\parallel} \times (i\mathbf{k}_{\parallel} \cdot i\mathbf{k}_{\parallel} \cdot \mathbf{O}^{(m)}) + \frac{1}{2}\partial_z \times (i\mathbf{k}_{\parallel} \cdot \partial_z \cdot \mathbf{O}^{(m)}) \\
&\quad + \frac{1}{2}\partial_z \times (\partial_z \cdot i\mathbf{k}_{\parallel} \cdot \mathbf{O}^{(m)}) \Big] \bigg\}.
\end{aligned}
\end{equation}

\section{Evolution of Field Discontinuities with Origin Translation}\label{sec:origin_translation}

To understand how the field discontinuities evolve as the origin moves, we can link the spatially varying boundary conditions at an arbitrary position $z$ directly to the physical reference origin at $z=0$.

By starting from the generalized extraction equations derived in the main text (Eq.~\ref{eq:extracted_jumps_substrate}), we first evaluate the field jumps exactly at the origin ($z=0$):
\begin{subequations}\label{eq:jumps_origin}
\begin{align}
    2\widehat{\Delta H}(0) &= \left( t_\text{tot} - e^{-jk d/2} \right) e^{jk d/2} + \left( r_\text{tot} + r_\text{s} t_\text{tot} e^{-jk d/2} \right), \\
    2\widehat{\Delta E}(0) &= \left( t_\text{tot} - e^{-jk d/2} \right) e^{jk d/2} - \left( r_\text{tot} + r_\text{s} t_\text{tot} e^{-jk d/2} \right).
\end{align}
\end{subequations}

By adding and subtracting these two equations, we can cleanly isolate the transmission and reflection terms as a function of the reference boundary conditions:
\begin{subequations}\label{eq:isolated_TR_origin}
\begin{align}
    \left( t_\text{tot} - e^{-jk d/2} \right) e^{jk d/2} &= \widehat{\Delta H}(0) + \widehat{\Delta E}(0), \\
    r_\text{tot} + r_\text{s} t_\text{tot} e^{-jk d/2} &= \widehat{\Delta H}(0) - \widehat{\Delta E}(0).
\end{align}
\end{subequations}

We now substitute these isolated terms back into the general $z$-dependent extraction formulas (Eq.~\ref{eq:extracted_jumps_substrate}). This eliminates the scattering parameters entirely, yielding a direct relationship between the boundaries at an arbitrary plane $z$ and those at $z=0$:
\begin{subequations}\label{eq:jumps_z_substituted}
\begin{align}
    2\widehat{\Delta H}(z) &= \big[ \widehat{\Delta H}(0) + \widehat{\Delta E}(0) \big] e^{jk z} + \big[ \widehat{\Delta H}(0) - \widehat{\Delta E}(0) \big] e^{-jk z}, \\
    2\widehat{\Delta E}(z) &= \big[ \widehat{\Delta H}(0) + \widehat{\Delta E}(0) \big] e^{jk z} - \big[ \widehat{\Delta H}(0) - \widehat{\Delta E}(0) \big] e^{-jk z}.
\end{align}
\end{subequations}

After simplification, we arrive at
\begin{equation}\label{eq:parity_rotation1}
    \begin{bmatrix}
        \widehat{\Delta H}(z) \\[1ex]
        \widehat{\Delta E}(z)
    \end{bmatrix}
    =
    \begin{bmatrix}
        \cos(k z) & j\sin(k z) \\
        j\sin(k z) & \cos(k z)
    \end{bmatrix}
    \begin{bmatrix}
        \widehat{\Delta H}(0) \\[1ex]
        \widehat{\Delta E}(0)
    \end{bmatrix}.
\end{equation}

\section{Multipolar variation under conicity}\label{sec:Multipole_conicity}

\begin{figure}[H]
    \centering
    \includegraphics[width=0.7\textwidth]{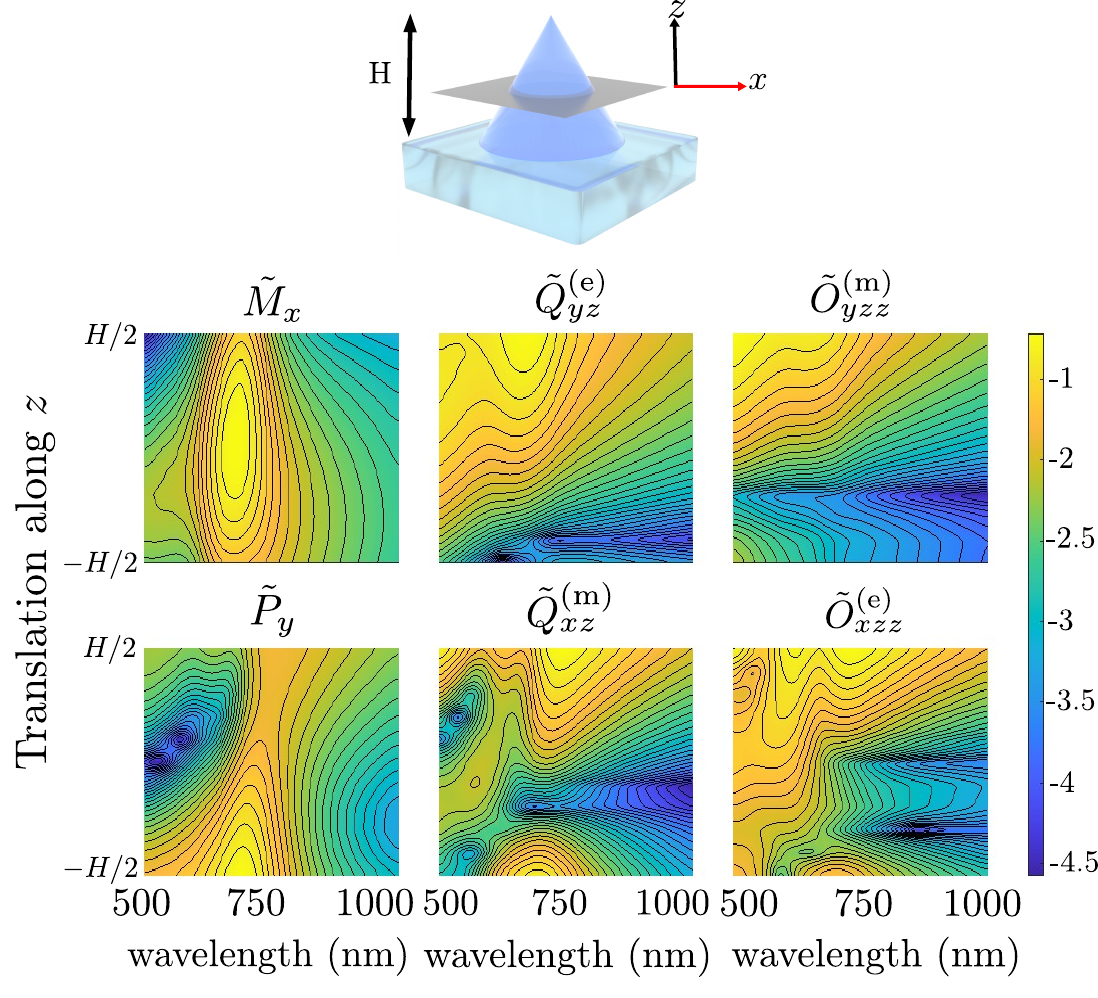}
    \caption{2D plot of the multipolar variations based on wavelength and translation for the cone represented in Figure}
    \label{fig_9:design}
\end{figure}

\begin{figure}[H]
    \centering
    \includegraphics[width=0.7\textwidth]{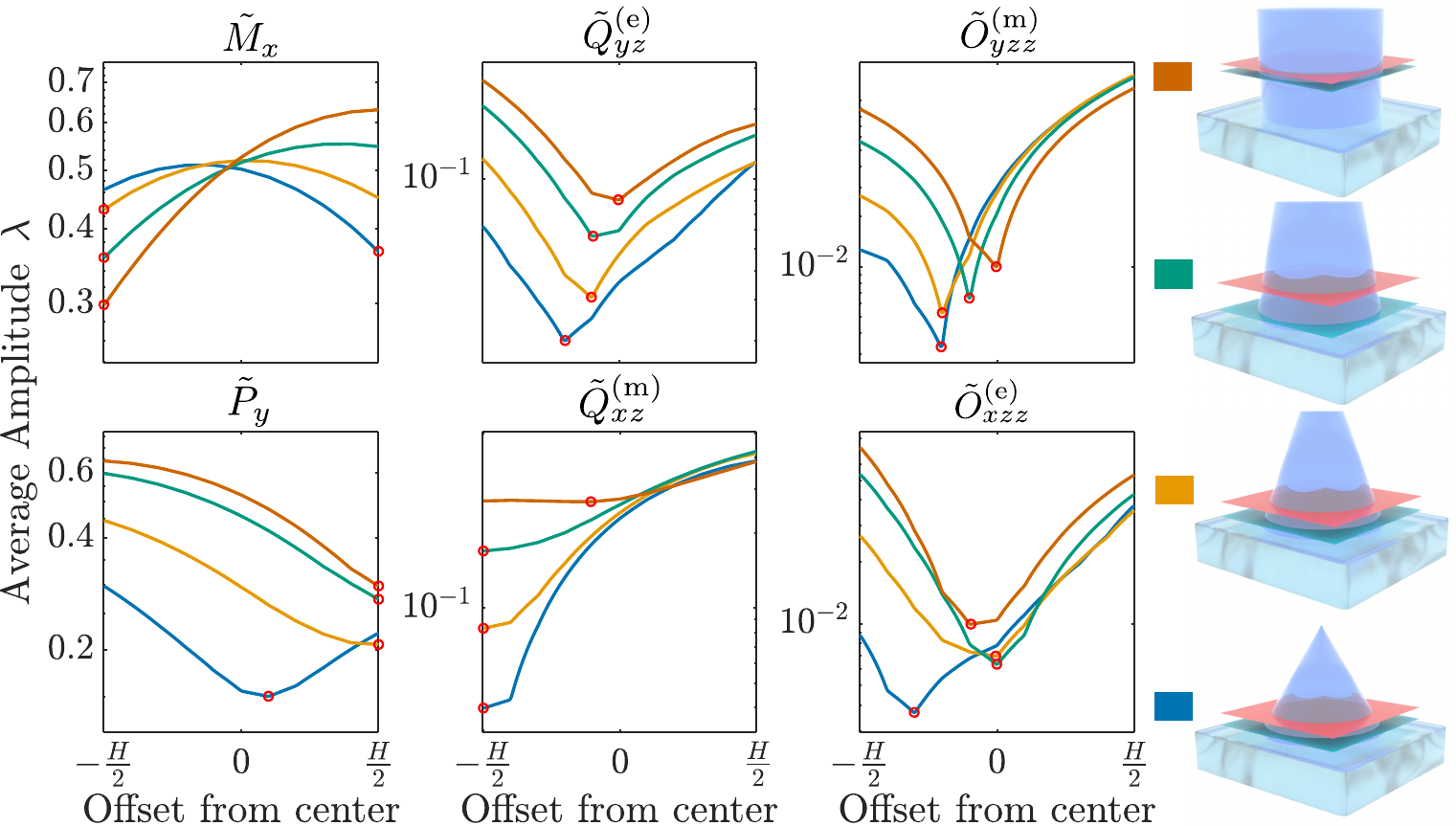}
    \caption{Variation of the multipolar moments under translation along ~$z$ for different levels of conicity. Each color in the corresponding multipole plots represents a distinct conicity level, as indicated in the side panel.}
    \label{fig_10:design}
\end{figure}

\section{Numerical errors when retrieving scattering parameters}\label{sec:Erroe}

\begin{figure}[H]
    \centering
    \includegraphics[width=0.9\textwidth]{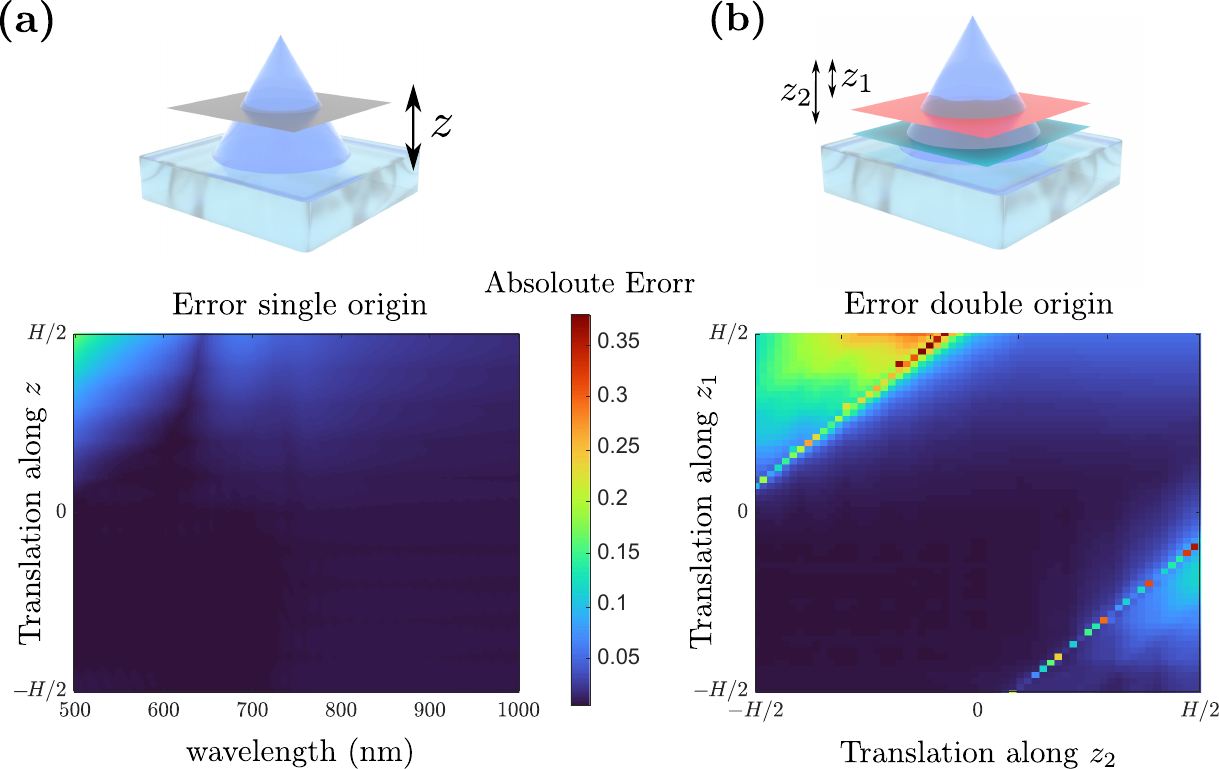}
    \caption{Numerical errors when retrieving scattering parameters. (a) Error for GSTCs evaluated at the same position. This error represents the contribution of multipole terms higher than octupoles in the non-zero regime. (b) Error for split-position GSTCs, calculated by separately translating the GSTC origins. The color map represents the average value of the scattering parameters over the wavelength, and the colored regions indicate the disallowed domains for the split GSTCs}
    \label{fig_3:design}
\end{figure}

Retrieving scattering parameters from multipole moments involves two primary sources of numerical errors. The first error source relates to the truncation of the multipole expansion, specifically, the number of multipoles required to accurately model the metasurface. This requirement depends on several factors, including the geometric complexity and symmetry of the structure, the electrical size of the object relative to the wavelength, and the material contrast. For cone studied in figure~\ref{fig_4:design}, we can find this error as figure~\ref{fig_3:design}.a. This figure shows retrieval scattering parameters from multipoles when we are placing our bouth boundary conditions at same place versus wavelength.

The second source of error, which comes from Equation~\ref{eq:RT_final_substrate}, arises from mathematical singularities. This occurs due to the degeneracy of the system described in Equation~\ref{eq:matrix_substrate_sep}, or when the denominator of Equation~\ref{eq:RT_final_substrate} approaches zero. This specific condition happens when the argument of the cosine function goes to zero:
\begin{equation}
    \cos\left[\frac{2\pi}{\lambda}(z_2 - z_1)\right] = 0
\end{equation}
Solving for the zeros of the cosine function reveals that this singularity occurs at specific path differences:
\begin{equation}
    z_2 - z_1 = \frac{(2m + 1)\lambda}{4}
\end{equation}
where $m$ is an integer ($0, \pm1, \pm2, \dots$). This indicates that the error peaks whenever the physical distance $(z_2 - z_1)$ corresponds to an odd multiple of a quarter .
 figure~\ref{fig_3:design}.b shows retrieval scattering parameters from ~\eqref{eq:RT_final_substrate} for two different offsetinf corespond to different parity boundary conditions which has been averaged over wavelength for both reflection and transmission.
 
\end{document}